\documentclass[aps,prl,longbibliography,twocolumn,nofootinbib, superscriptaddress]{revtex4-2}
\usepackage{amsmath}
\usepackage{amssymb}
\usepackage{graphicx}
\usepackage{comment}
\usepackage{soul,xcolor}
\usepackage{bbold}
\usepackage[colorlinks=true ,urlcolor=blue,urlbordercolor={0 1 1}]{hyperref}
\usepackage{tikz}
\usepackage{braket}
\usepackage[english]{babel}
\usepackage{graphicx}
\usepackage{lipsum}
\usepackage{xr}
\usepackage{ulem}

\usepackage[version=4]{mhchem}

\newcommand{\<}{\langle}

\renewcommand{\>}{\rangle}
\renewcommand{\(}{\left(}
\renewcommand{\)}{\right)}

\renewcommand{\v}[1]{\mathbf{#1}} 

\renewcommand{\d}{\partial}

\newcommand{\eps}{\epsilon}
\renewcommand{\t}{\tau}
\renewcommand{\b}{\bar}
\newcommand{\ba}{\begin{align}}
\newcommand{\ea}{\end{align}}
\renewcommand{\pl}{\parallel}
\newcommand{\pp}{\perp}

\newcommand{\tb}[1]{ \textcolor{blue} }

\renewcommand{\d}{^{\dagger}}
\renewcommand{\vec}[1]{{\boldsymbol #1}}

\begin{document}

\title{Non-Fermi Liquids from Subsystem Symmetry Breaking in van der Waals Multilayers}

\author{Archisman Panigrahi}
\affiliation{Department of Physics, Massachusetts Institute of Technology, Cambridge, Massachusetts 02139, USA}

\author{Ajesh Kumar}\email{ajeshk@mit.edu}
\affiliation{Department of Physics, Massachusetts Institute of Technology, Cambridge, Massachusetts 02139, USA}

\date{\today}
\begin{abstract}
We investigate the spontaneous breaking of subsystem symmetry in a stack of two-dimensional Fermi liquid metals, each maintaining a subsystem number conservation symmetry, driven by interlayer exciton condensation. 
The resulting Goldstone modes in this broken symmetry phase couple to the quasiparticle current perpendicular to the layers. This coupling, which remains non-zero for small momentum transfers, leads to the emergence of a three-dimensional anisotropic marginal Fermi liquid state when the number of layers is sufficiently large. We propose a possible experimental realization of this phenomenon in two-dimensional multilayer van der Waals heterostructures. 
Using self-consistent mean-field calculations, we characterize the subsystem symmetry-broken metallic state and examine the effects of fluctuations on its physical properties within the random phase approximation. We find that these fluctuations produce additional logarithmic enhancements to the specific heat at low temperature, specifically $C\sim T [\log(1/T)]^2$.
\end{abstract}
\maketitle

Non-Fermi liquids (NFL) are a class of metallic phases of matter that deviate from the paradigmatic Landau Fermi liquid theory. 
A significant amount of theoretical and experimental effort continues to be devoted to studying these complex, strongly-coupled phases and their experimental realizations~\cite{lohneysen2007fermi,lee2018recent, CHUBUKOV2020168142, NAYAK1994359, SSLee2009, Metlitski2010a, Metlitski2010b, Mandal2015, Holder2015}.
They are typically believed to emerge in strongly correlated materials such as cuprates and pnictides when a bosonic order parameter becomes quantum critical~\cite{lee2006doping, paschen2021quantum, taillefer2010scattering}. 
A recent addition to the catalog of strongly correlated materials are two-dimensional van der Waals materials,
which are heterostructures formed by stacking single material layers on top of each other~\cite{andrei2021marvels,nuckolls2024microscopic,mak2022semiconductor,cao2018unconventional,zhou2021superconductivity,kennes2021moire,cai2023signatures,Mandal2022Valley-pol-nematic,zeng2023thermodynamic,lu2024fractional, wang2019evidence, shi2022bilayer, zhang2022correlated, Joe2024, liu2024optical, nguyen2025quantumoscillationsdipolarexcitonic}. 
These heterostructures, created by stacking individual material layers, offer remarkable flexibility through gate-voltage tunability and diverse layer combinations, each yielding distinct physical properties.
This versatility raises the intriguing possibility of these materials facilitating novel pathways to the realization of NFLs.

\begin{figure}[t]
    \centering
    \includegraphics[width=0.28\textwidth]{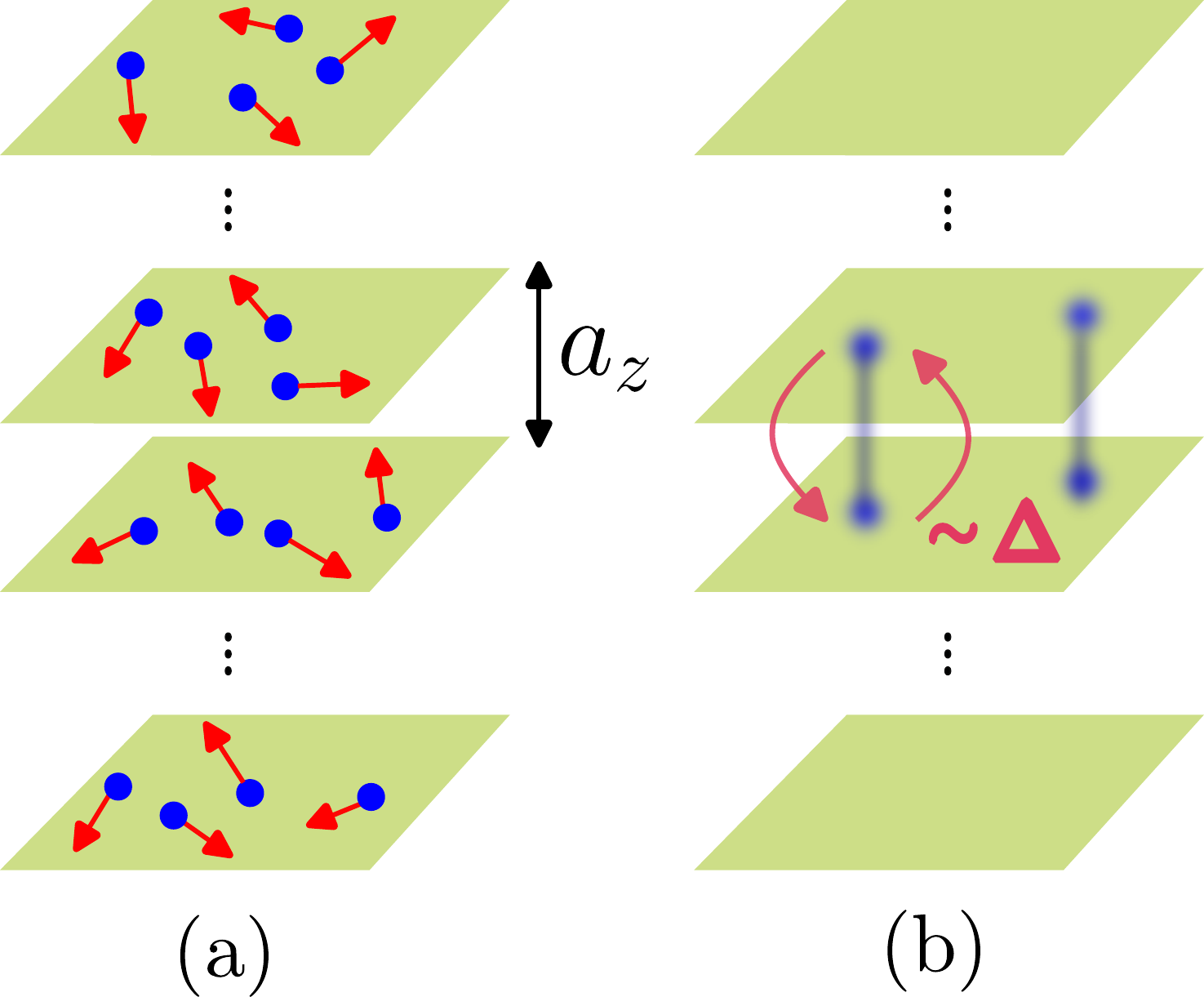}
    \caption{(a) Stack of two-dimensional Fermi liquid metals with subsystem symmetry and long-lived quasiparticles. (b) Inter-layer coherent, or subsystem symmetry broken, three-dimensional non-Fermi liquid metal with short-lived single-particle excitations.}
    \label{fig:figure1}
\end{figure}

In metallic states that spontaneously break continuous symmetries, the coupling of electrons to Goldstone modes can, in certain cases, result in NFL behavior~\cite{oganesyan2001quantum, xu2010quantum,watanabe2014criterion,bahri2015stable,kumar2020fractal,lake2023nfl,anakru2023nfl, Ghorai2023PhysRevB.108.045117}.
In scenarios where the broken symmetry generators affect only internal flavor degrees of freedom—such as in an XY ferromagnetic metal—Goldstone modes decouple from the electrons at low energies, thus maintaining the stability of the Fermi liquid~\cite{watanabe2014criterion}. 
Let us consider a stack of multiple two-dimensional metallic layers, where electrons in each layer can interact with those in the adjacent layers. Each layer possesses a subsystem particle number ($n_i$ for the $i$-th layer) conservation symmetry, and there is a discrete translational symmetry perpendicular to the layers. The relative number ($n_i - n_{i+1}$) symmetry can be spontaneously broken through interlayer exciton condensation, while the total electron number conservation remains intact. Owing to the fact that the electrons reside in spatially separated layers, the subsystem symmetry is not a true internal symmetry. We show that the coupling between electrons and the Goldstone modes of the broken subsystem symmetry remains finite even at arbitrarily small momentum transfer, and this strong coupling gives rise to a non-Fermi liquid state.

It is important to note that subsystem symmetry breaking at zero temperature necessitates that the spatial dimension of the subsystem where the symmetry generators act non-trivially is at least two. 
In lower-dimensional subsystems, strong quantum fluctuations will restore the system to a symmetry-preserving state~\cite{batista2005generalized}.
Field theories with subsystem symmetries and their spontaneous breaking have also been considered recently in the context of fractonic systems (for a sample of papers, see Refs.~\cite{nandkishore2019fractons,pretko2020fracton,gromov2024fracton,seiberg2021exotic,distler2022spontaneously,han2022nfl}).

Subsystem number conservation symmetry can be realized in $2D$ materials when monolayers are separated by an insulating barrier such as boron nitride~\cite{eisenstein2004bose,eisenstein2014exciton}, or when the layers are arranged with a large-angle twist~\cite{kim2018spin}, significantly suppressing interlayer electron tunneling. 
In the bilayer setting with suppressed tunneling, several studies have reported interlayer coherence in both electron-electron and electron-hole systems~\cite{wang2019evidence, shi2022bilayer, liu2024optical}. Building on these findings, we extend the analysis to multilayer setups.
We begin by analyzing a microscopic model of such a multi-layer system. Using self-consistent mean-field calculations, we examine the phase where subsystem symmetry is spontaneously broken.
We then derive a low-energy model of a $3D$ Fermi liquid coupled to Goldstone mode fluctuations, demonstrating that this interaction leads to a marginal Fermi liquid state, whose properties we subsequently analyze.






\textit{Mean-field theory}: Consider a system of $N$ two-dimensional metallic layers. In the layer labelled $i$, electrons with in-plane momentum $\v k_{2D}$ have a dispersion $\eps_{{k_{2D}}}$, and are described by the fermionic annihilation operators $f_{\v k_{2D},i}$.
We assume that the layers are identical so that the system has discrete translational symmetry in the out-of-plane ($z$) direction.
The microscopic Hamiltonian includes 
the kinetic energy of the electrons, as well as electronic-interactions both within the same layer and between adjacent layers,
\begin{equation}\label{eq:microscopic-Hamiltonian}
\begin{aligned}
    H&= \sum_{\v k_{2D},i}(\epsilon_{k_{2D}} -\mu) f^\dagger_{\v k_{2D},i} f_{\v k_{2D},i} \\
    &+ \frac{1}{2A} \sum_{\v k_{2D}, \v k'_{2D}, \v q,i} \tilde{V}_0(\v q) f^\dagger_{\v k'_{2D} + \v q,i} f^\dagger_{\v k_{2D} - \v q,i} f_{\v k_{2D},i} f_{\v k'_{2D},i} \\
    &+  \frac{1}{A} \sum_{\v k_{2D}, \v k'_{2D}, \v q, i} \tilde{V}_{int}(\v q) f^\dagger_{\v k'_{2D} + \v q,i+1} f^\dagger_{\v k_{2D} - \v q,i} f_{\v k_{2D},i} f_{\v k'_{2D},i+1}
\end{aligned}
\end{equation}
Here, $A$ is the system area.
$\tilde{V}_0(\v q) = \frac{e^2}{2\epsilon_0 \epsilon_r q}$, and $\tilde{V}_{int}(\v q) = \frac{e^2}{2\epsilon_0 \epsilon_r q}e^{-qa_z}$ are respectively the intra and inter-layer Coulomb interactions, and $a_z$ is the inter-layer separation.
We utilize a parabolic dispersion, $\epsilon_k = \frac{\hbar^2 k^2}{2 m^*}$, with $m^* = 0.07 m_e$, the effective band mass in Gallium Arsenide~\cite{sze2021physics} 
, chosen here as a representative example. We also set the static dielectric constant $\epsilon_r = 12.5$ as observed in Gallium Arsenide.

The number of fermions in each layer $n_i$ is separately conserved, implying that a $U(1)$ symmetry exists for each layer. 
Since we will be interested in breaking this down to a global number-conservation symmetry, we decompose the conserved quantities into charge differences between adjacent layers and the total charge: $n_{i}-n_{i+1}$ and $n_t=\sum_i n_i$. The corresponding symmetry group is $U(1)^1 \times U(1)^2 \ldots U(1)^{N-1} \times U(1)^t$. 
Let us now consider the spontaneous breaking of the interlayer charge conservation symmetries and keep only $U(1)^t$ unbroken. This state is defined by the interlayer exciton order parameter $\chi_i(\v k_{2D}) \equiv \<f\d_{\v k_{2D},i+1} f_{\v k_{2D},i} \> \neq 0$. 

\begin{figure}[t]
    \centering
    \includegraphics[width=0.49\textwidth]{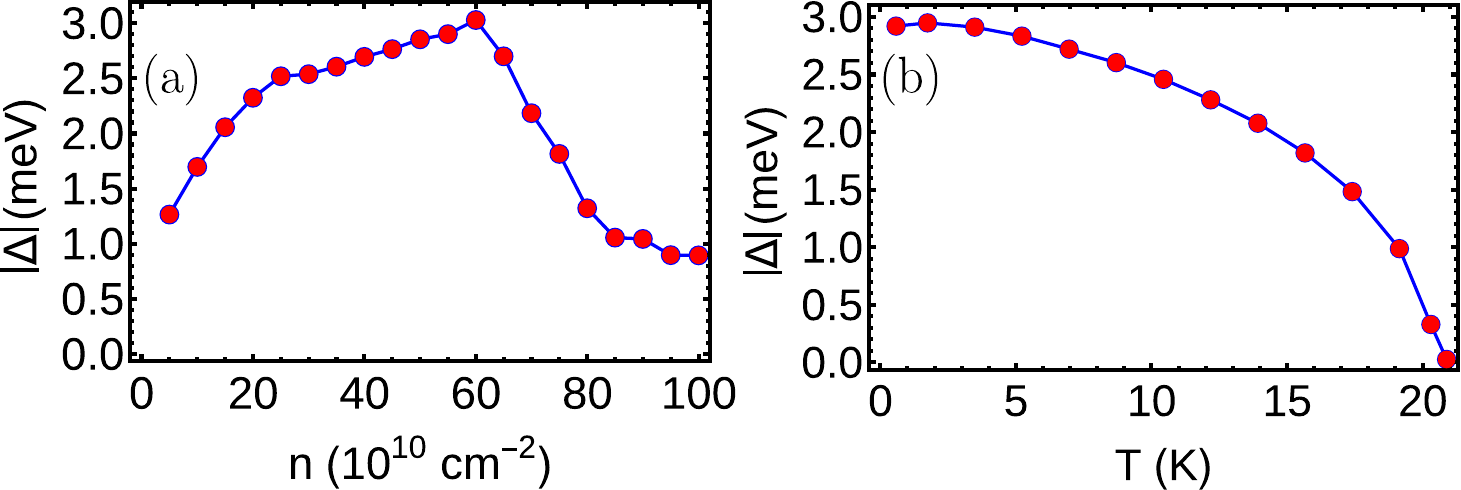}
    \caption{(a) Dependence of the maximum value of $|\Delta|$ (over all momenta), as a function of electron density at a fixed temperature ($T = 0.05 \text{ meV}/k_B \approx 0.58$ K) for a system of 10 layers, with inter-layer distance $a_z = 3$ nm. (b) Dependence of the maximum value of $|\Delta|$ as a function of temperature, at a fixed density, $n = 55 \times 10^{10}/$cm$^2$. For both the plots, we set $m^* = 0.07 m_e$ and $\epsilon_r = 12.5$.}
    \label{fig:figure2}
\end{figure}

To simplify the analysis, we omit the in-plane interaction as it does not contribute to excitonic ordering. Similarly, the Hartree terms, which merely renormalize the chemical potential, are also disregarded. 
While intralayer interactions can, in principle, drive further instabilities such as charge density wave ordering, we neglect these effects motivated by recent experiments that observe only interlayer exciton ordering~\cite{wang2019evidence, liu2024optical}.
 In the presence of such charge density wave ordering, the Fermi surface in each layer becomes reconstructed, except at special fillings where the CDW can fully gap the Fermi surface. The subsequent analysis can then be applied to the resulting metallic state, and our main conclusions remain unaffected.

Within the mean-field approximation, the effective Hamiltonian features interaction-induced hopping between adjacent layers \cite{zheng1997exchange,wu2015theory,Zhu2023TwoDimensionCoherence}, and it can be written as,
\begin{equation}\label{eq:mean-field-hamiltonian}
\begin{aligned}
    H_{MF} = &\sum_{\v k_{2D},i} (\epsilon_{k_{2D}}-\mu) f^\dagger_{\v k_{2D},i} f_{\v k_{2D},i} \\
        &- \sum_{\v k_{2D},i} \left(\Delta(\v k_{2D},i) f^\dagger_{\v k_{2D},i} f_{\v k_{2D},i+1} + h.c.\right)
\end{aligned}
\end{equation}
where the exciton gap function $\Delta$ is defined as,
\begin{equation}\label{eq:delta-definition}
\begin{aligned}
    \Delta(\v k_{2D},i) &= \frac{1}{A} \sum_{\v k'_{2D}} \tilde{V}_{int} (\v  k_{2D} - \v k'_{2D}) \<f^\dagger_{\v k'_{2D},i+1} f_{\v k'_{2D},i} \>
\end{aligned}
\end{equation}
It is interesting to note that the mean-field Hamiltonian in Eq.\eqref{eq:mean-field-hamiltonian} effectively describes a tight-binding model along the $z$ direction, implying that it can transport current in this direction. However, the original Hamiltonian in Eq.\eqref{eq:microscopic-Hamiltonian} did not allow any electron transport in the out-of-plane direction. While the system is apparently a $3D$ metal at the mean-field level, it is expected that the fluctuations of the Goldstone boson will destroy the conductivity along $z$ direction, while preserving the $3D$ Fermi surface. A similar scenario occurs in mean-field Hamiltonians of dipole conserving systems under uniform electric field \cite{lake2023nfl}. 
We describe transport properties in greater detail in a later section. 

To solve the mean-field problem, we employ periodic boundary condition (for the simplicity of numerical calculation and since our interest in this section is in the bulk mean-field electronic bandstructure) in the $z$ direction, such that $k_z \in \frac{2\pi}{N a_z}\times \{1,2,3,\cdots, N \}$ is a good quantum number. Let us assume that $\langle f^\dagger_{\v k,i+1} f_{\v k,i} \rangle$ is uniform in the $z$ direction, i.e., it does not depend on the layer index $i$. Then, $\Delta(\v k_{2D},i)$ becomes $i$ independent.
The energy eigenvalues of the mean-field Hamiltonian are,
\begin{equation}\label{eq:energy-eigenvalues}
    E_{\v {k}_{2D}, k_z} = \frac{\hbar^2 k_{2D}^2}{2 m^*} - 2 \Delta(\v k_{2D}) \cos(k_z a_z)
\end{equation}
By allowing inter-layer hopping, the exciton ordering enables a dispersion along $k_z$ for the fermions\footnote{We note that $\Delta$ can admit a layer-dependent phase factor since the order parameter is independent for every pair of layers. However, this phase can be eliminated by a redefinition of the fermion degrees of freedom.}.
For this dispersion to result in a $3D$ metallic behavior, a sufficiently large $N$ is required, so that the finite energy gap between states becomes unresolvable. 
Linearizing the dispersion near the Fermi surface along $k_z$, the smallest excitation energy is $E_g = 2\Delta (\delta k_z) a_z$, where $(\delta k_z) a_z = 2\pi/N$ is the smallest change in $k_z$, where $\Delta$ is evaluated at the Fermi surface.
If the probe energy scale $E$ significantly exceeds this energy gap, $E > E_g$,  the system exhibits characteristics of a $3D$ metal with gapless excitations near a Fermi surface.
Additionally, the NFL physics that we describe later relies on interlayer coherence, so energy scale must remain below the maximum value of $\Delta$, which happens for $N > 4 \pi \approx 12.57$. 
Consequently, for $N \lesssim 12$, the system will behave as a $2D$ Fermi liquid metal and the non-Fermi liquid phenomena explored in this study would likely not be detectable.

Rewriting in terms of momentum $k_z$, Eq.\eqref{eq:delta-definition} simplifies,
\begin{equation}\label{eq:delta-definition-periodic}
\begin{aligned}
    \Delta(\v k_{2D}) = \frac{1}{A N} \sum_{\v k'_{2D}, k_z} \tilde{V}_{int}&(\v k_{2D} - \v k'_{2D}) \cos(k_z a_z) \\
    &\times \<f^\dagger_{\v k'_{2D},k_z} f_{\v k'_{2D},k_z} \>,
\end{aligned}
\end{equation}
where the occupation numbers are given by the Fermi distribution function,
\begin{equation}\label{eq:fermi-function}
    \<f^\dagger_{\v k_{2D},k_z} f_{\v k_{2D},k_z} \> = \frac{1}{e^{(E_{\v k_{2D},k_z} - \mu)/k_B T} + 1}.
\end{equation}
We iteratively solve the three equations \eqref{eq:energy-eigenvalues}, \eqref{eq:delta-definition-periodic} and \eqref{eq:fermi-function} self-consistently to obtain the equilibrium value of the excitonic condensate $\Delta(\v k_{2D})$ under a fixed electronic density. The Julia program used to self-consistently calculate mean-field solution is available at Ref.\cite{panigrahi_subsystem_symmetry_breaking_2024_codes}. At a any fixed density, we create a $51 \times 51$ rectangular grid in momentum space, with the grid boundary extending to approximately twice the Fermi momentum. We have plotted the variation of the maximum absolute value of $\Delta(\v k_{2D})$ as a function of electronic density in Fig.\ref{fig:figure2}(a). We find that at small densities, $|\Delta|_{\text{max}}$ increases with rising density. Eventually, it reaches a maximum, and then falls off as density increases further. 
At small densities, very few bands (or equivalently $k_z$-states) are filled, and they contribute to $\Delta$ constructively (i.e., all non-zero terms in the right hand side of Eq.\eqref{eq:delta-definition-periodic} have the same sign). But, at higher densities, all the bands become partially filled, and the filling of the bands with higher energy effectively cancel the contribution of the low-energy bands (different $k_z$ contributions in the right hand side of Eq.\eqref{eq:delta-definition-periodic} have different signs). As a result, beyond a certain density (where the bands are `half-filled', i.e., when the Fermi wave-vector along $z$ is  $k_{F,z} = \pi/(2 a_z)$), the order parameter begins to decrease. For the parameters mentioned in the figure caption, the maximum value of $\Delta$ is obtained near density $55 \times 10^{10}/$cm$^2$. At this density, we obtain the self-consistent maximum value of $\Delta$ and plot it as a function of temperature in Fig.\ref{fig:figure2}(b). Under the mean-field approximation (that is known to overestimate the critical temperature), we find that a continuous phase transition occurs at some temperature $T_c$ between 20.3 K to 20.9 K.

The spontaneous symmetry breaking discussed here leads to a Goldstone mode for every pair of nearest-neighbor layers. We next analyze how these modes couple to the mean-field fermionic excitations.

\textit{Fermion-Goldstone mode coupling}: Let us introduce fluctuations about the mean-field state --- $\Delta_i(\v r)$ denotes the exciton gap function between adjacent layers $i$ and $(i+1)$ --- here we employ open boundary conditions and $i$ takes values from 1 to $(N-1)$.
We consider a mean-field gap function that is uniform in $\v k_{2D}$.
This assumption, valid for sufficiently short-range interactions, captures the qualitative essence of the symmetry-broken phase.
This results in the following term in the fluctuation action,
\begin{align}
    S_{\Delta-f} = -\int d \t \sum_{i=1}^{N-1} \sum_{\v r}  \Delta_i(\v r_{2D}, \t) \b f_{i}(\v r_{2D},\t) f_{i+1}(\v r_{2D},\t) + \text{h.c.},
\end{align}
where $\v r_{2D}$ is a $2D$ coordinate in the plane of the layers.
Focusing on the low-energy phase fluctuations of $\Delta_i(\v r_{2D},\t) = \Delta_0 e^{i\phi_i(\v r,\t)}$, we obtain, $S_{\Delta-f} = -\Delta_0 \int d \t \sum_{i=1}^{N-1}\sum_{\v r_{2D}} i\phi_{i}(\v r_{2D},\t) \b f_{i}(\v r_{2D},\t) f_{i+1}(\v r_{2D},\t) + \text{h.c.}$, which in the continuum yields
\begin{align}
    S_{\Delta-f} = \int_{\v r, \t} j_z(\v r,\t) \frac{\phi(\v r,\t)}{e},
\end{align}
where $j_z = -i e \b{f} \partial_z f/(2m_z) + \text{h.c.}$, is the fermion number current along $z$, $m_z^{-1} \equiv 2\Delta_0 a_z^2$, $\v r$ is a $3D$ spatial coordinate, and $\int_{\v r,\t} \equiv \int d^3r d\t$. 
In going from lattice sum to spatial integral, we have rescaled the current operator by the volume of a unit cell and $\phi$ by $a_z$.
Here we emphasize that the qualitative form of this fermion-Goldstone mode coupling does not rely on details of the electron dispersion. In particular, it applies even if fermions develop instabilities in the intralayer channel that modify the quasiparticle dispersion.

The spontaneous breaking of the $N-1$ interlayer charge conservation symmetries results in $N-1$ Goldstone modes: $S_{\phi_i} = \rho_s a_z \int_{\v r_{2D},\t} (\partial_{\t} \phi_i(\v r_{2D},\t))^2/c^2 + (\partial_{\v r_{2D}} \phi_i(\v r_{2D},\t)^2)$.
It can be readily verified that terms like $(\phi_i(\v r_{2D}) - \phi_{i+1}(\v r_{2D}))^2$ give rise to gapped modes, and hence, are not allowed in order to have $N-1$ gapless modes. Consequently, the continuum action cannot have terms like $(\partial_z \phi(\v r))^2$. In other words, the effective low-energy action should have the same subsystem symmetry as in the parent Hamiltonian, which implies that spatial fluctuations of the form: $\phi_i(\v r_{2D}, \tau) \rightarrow \phi_i(\v r_{2D}, \tau) + g_i$ for some arbitrary numbers $g_i$ (in the continuum limit,  $\phi(\v r_{3D}, \tau) \rightarrow \phi(\v r_{3D}, \tau) + g(z)$), cost zero energy.
Since terms such as $(\partial_z \phi)^2$ are not allowed by the subsystem symmetry, the $\phi$ field has a flat dispersion along $z$.
All the Goldstone modes have the same stiffness $\rho_s$ and speed $c$, due to translational symmetry along $z$. 
In the continuum, when summed over all the pairs of adjacent layers labelled $i = 1,2,\cdots (N-1)$,
\begin{align}
    S_{\phi} = \rho_s \int_{\v r,\t} \frac{1}{c^2}\(\partial_{\t} \phi(\v r,\t)\)^2 + \(\partial_{\v r_{2D}} \phi(\v r,\t)\)^2
    \label{eq:phase_fluc}
\end{align}
Additionally, we exclude the effect of vortices in $\phi$ which, in the presence of strong disorder, could become pinned and disrupt interlayer coherence. However, we assume a sufficiently clean setup and neglect this effect in our analysis.

Our next step is to construct a low-energy model of the $3D$ metal that we can use to calculate its physical properties. Consider fermions near a patch on the Fermi surface at an angle $\theta$ measured relative to the $z$-direction, with Fermi momentum $\v k_F(\theta)$. 
Due to the system's rotational symmetry about $z$, an additional azimuthal patch label is unnecessary.
The low-energy fermionic action at this patch is: 
\begin{align}\label{eq:just-Fermion-action}
    S_f = \int_{\v k, \omega} \b{f}(\v k,\omega) \(i\omega - v_F(\theta) k_{\pp} - \frac{k_\pl^2}{2m(\theta)} \) f(\v k,\omega)
\end{align}
where Fermi velocity $v_F$, mass $m$ are patch-dependent parameters, $k_{\pp}, k_{\pl}$ are respectively the momenta measured perpendicular and tangential relative to the patch and $\int_{\v k, \omega} \equiv \int \frac{d^3k d\omega}{(2\pi)^4}$. The coupling to Goldstone modes is given by the matrix elements of the current operator with fermionic states at momentum $\v k$ and $\v k + \v q$: $\<\v k|j_z| \v k + \v q\> = \Delta_0 a_z^2 (2k_z + q_z)$. By taking the limit $\v q \rightarrow 0$, $\v k \approx \v k_F$, and keeping only the leading $\v q$-independent term, we obtain
\begin{align}
    S_{\phi-f} = \frac{k_{F,z}(\theta)}{m_z} \int_{\v k, \v q, \omega, \Omega} \phi(\v q,\Omega) \b{f}(\v k + \v q,\Omega+\omega) f(\v k,\omega)
\end{align}
where ${k_{F,z}(\theta)} = \v k_F(\theta) \cdot \hat{z}$. 
In the following analysis, we will incorporate the $a_z$ factor into $k_F$.

\textit{Non-Fermi Liquid properties}: Having constructed the low-energy model $S = S_f + S_{\phi-f} + S_{\phi}$, we now calculate various physical properties of the resulting state.
Since our focus is on the universal properties of the phase, in some parts of the calculations we will assume a spherical Fermi surface, and disregard the $\theta$-dependence of $v_F$, $k_F$ and $m$. 

\textit{A. Fermion self-energy}:
\begin{figure}[t]
    \centering
    \includegraphics[width=0.49\textwidth]{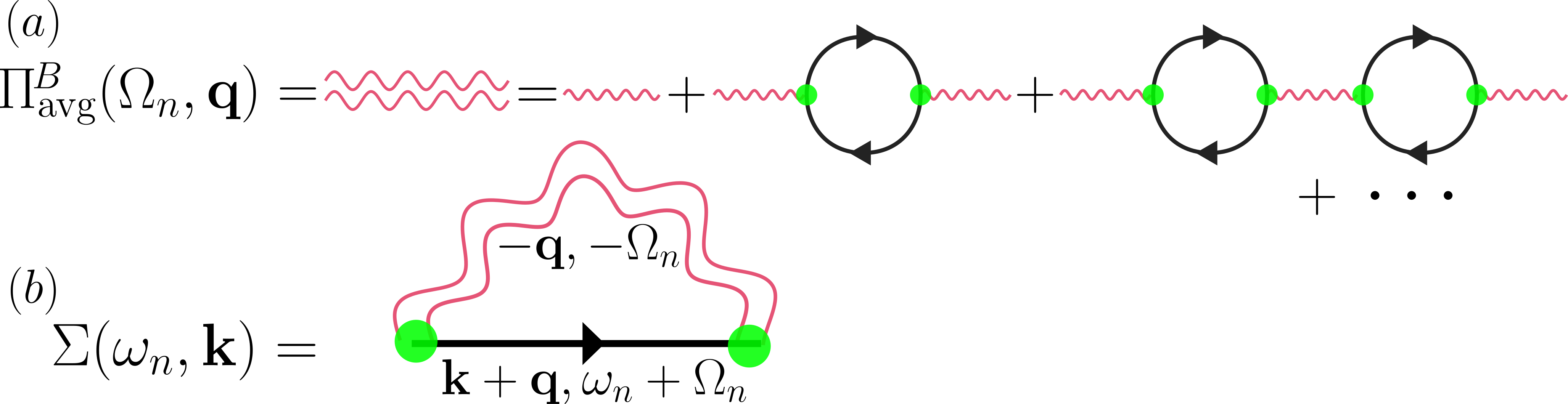}
    \caption{Feynman diagrams at the level of RPA for (a) Bosonic self-energy and (b) Fermionic self-energy.}
    \label{fig:feynman-diagrams}
\end{figure}
Let us begin with the computation of the one-loop fermion self-energy, whose imaginary part corresponds to the inverse-lifetime of a quasiparticle. The standard procedure involves first calculating the boson self-energy, followed by using the corrected boson propagator within the random phase approximation (RPA) to compute the fermion self-energy~\cite{sachdev1999quantum}. The boson self-energy is calculated, by averaging over all Fermi surface patches, to give the standard Landau-overdamped form, with an additional anisotropy factor:
\begin{align}
    \Pi^B_{\text{av}}(\Omega, \v q) = \frac{\Lambda k_F^2 \Delta_0^2 m}{4 v_F}\frac{|\Omega|}{q} \(1+\frac{q_z^2}{q^2}\) \equiv \gamma \frac{|\Omega|}{q}\(1+\frac{q_z^2}{q^2}\),
\end{align}
where $\Lambda$ is a UV momentum cutoff. 
Using the modified boson propagator, we calculate the fermion self-energy:
\begin{align}\label{eq:fermion-self-energy-result}
\Sigma(&\omega_n,\v k ) =  4\Delta_0^2 k_{F,z}(\theta)^2 \int \frac{dq_{\pp} d^2q_{\pl} d \Omega}{(2\pi)^4} \nonumber \\
&\frac{1}{i(\omega_n + \Omega) - v_F q_{\pp} - \frac{q_\pl^2}{2m}} \frac{1}{\rho_s q_{\pl}^2(1 - (\hat{q}_{\pl}  \cdot \hat{z})^2) + \Pi^B_{\text{av}}(\Omega,\v q_{\parallel})} \nonumber \\
&= i \frac{2\Delta_0^2 k_F^2 |\cos\theta|}{3\pi^2\rho_s v_F} \text{sgn}(\omega_n)|\omega_n| \log\left(\frac{\rho_s \Lambda^3}{\gamma |\omega_n|}\right)
\end{align}

Therefore, near the Fermi surface, the lifetime of the quasiparticles scales as $\tau(\omega) \sim \frac{1}{|\omega| \log(1/|\omega|)}$ instead of the $1/\omega^2$ behavior found in regular Fermi liquids.
The resulting metallic state is an anisotropic marginal Fermi liquid, characterized by short-lived single-particle excitations near the Fermi surface.
In particular, the prefactor of the self-energy vanishes for fermions in the $k_z = 0$ plane, leading to long-lived quasiparticles in this region. In contrast, marginal Fermi liquid behavior dominates across the rest of the Fermi surface.

\textit{B. Specific heat}: A frequently discussed feature of NFLs is the enhancement of specific heat at low temperatures, deviating from the linear dependence characteristic of Fermi liquids. 
The specific heat $C$ of an isotropic marginal Fermi liquid, such as a Fermi surface coupled to a fluctuating gauge field in three dimensions, exhibits a logarithmic enhancement, with $C/T \sim \log(1/T)$~\cite{hartnoll2022colloquium,senthil2004weak}.
Here we examine how the system anisotropy in the present context modifies this result.

The free energy within the random phase approximation (RPA) is given by
\begin{equation}
    \begin{aligned}
    \delta F(T)
    &=\frac{T}{2} \sum_{\Omega_n}\int \frac{d^3 q}{(2\pi)^3} \log \( \rho_s(q_x^2 + q_y^2) + \gamma \frac{|\Omega_n|}{q}\)\\
    &\qquad\qquad\qquad\qquad\qquad\qquad\qquad - F(T=0)\\
    &\sim - T^2 (\log(1/T))^2,
    \end{aligned}
\end{equation}
where $\Omega_n$ are bosonic Matsubara frequencies,
and the anisotropy factor $1+q_z^2/q^2$ has been omitted, as it remains finite and does not influence universal scaling properties.
We evaluate the above integral to find that the absence of dispersion along $z$ for the Goldstone mode leads to an additional enhancement of the specific heat (for details of the calculation, see appendix B in supplementary material\cite{supp}). 
Specifically, we find that specific heat $C \sim T \(\log (1/T)\)^2$.


\textit{C. Electronic transport}:
In the presence of an external electromagnetic vector potential $\v A$, the microscopic Hamiltonian in Eq.\eqref{eq:microscopic-Hamiltonian} gets modified via the standard minimal coupling of the $x$ and $y$ components of $\v A$ to the fermions. However, the $z$-component does not couple to the fermions due to the absence of any interlayer hopping, i.e., the microscopic current operator along $z$ is identically zero.
This implies that the system does not respond to any electric field along $z$, or any magnetic fields in the $x$-$y$ plane, regardless of the phase. 
In particular, there cannot be any electronic transport in the $z$ direction: $\sigma_{zz}(\omega, \v q) = 0$, for any frequency and momentum. 

We can however ask how in-plane transport gets affected due to fermion scattering off phase fluctuations $\phi$. 
We present calculations for their effect on the transport scattering rate, while neglecting both disorder and Umklapp scattering—factors essential for achieving a finite resistivity
(see appendix C in supplementary material for details \cite{supp}).
This standard treatment should be understood as providing finite-temperature corrections to a zero-temperature resistivity arising from disorder or Umklapp scattering~\cite{watanabe2014skx, lee1992gauge}.
Our calculations give a resistivity scaling $\rho \sim T^{5/3}$ due to scattering off Landau-overdamped phase fluctuations, which crosses over to $\rho \sim T^{3}$ at higher temperatures prior to the onset of Landau-damping.  

The discussion so far neglected interlayer tunneling, which explicitly breaks the subsystem symmetry. While it can can be strongly suppressed in experiments
, let us describe the consequence of having a small tunneling $t_e$. 
The gaplessness of the Goldstone modes is no longer protected and they develop a gap set by $t_e$. The NFL physics discussed in this letter would then apply only when the probe energy scale is $E > t_e$, at which point the Goldstone mode propagator takes the Landau-overdamped form. For $E<t_e$, the gap dominates and Fermi liquid behavior will be recovered (for details of the calculation, see appendix D in supplementary material\cite{supp}).

\textit{Discussion}:
In conclusion, we propose that interlayer exciton condensation in multilayer heterostructures with conserved electron number in each layer, can give rise to NFL behavior. Using a realistic microscopic model, we perform mean-field calculations to characterize the resulting symmetry-broken metallic state. 
We find a range of densities where interlayer coherence is achieved for temperatures $T \lesssim 20$ K within our mean-field calculations that neglect screening effects, which could suppress $T_c$~\cite{KharitonovScreening2008}. However, such large values of $T_c$ have been reported in recent experiments \cite{wang2019evidence, liu2024optical}.
This NFL phenomenon may also manifest in alternating electron-like and hole-like layers.
We highlight key physical properties that could serve as experimental signatures of the NFL phase.

Our work presents several promising avenues for future research. Since NFL behavior is most pronounced in two dimensions, one intriguing direction would be to explore subsystem symmetry breaking in a $2D$ system composed of $1D$ subsystems. While spontaneous exciton ordering is forbidden in purely $1D$ systems, it is possible to develop quasi-long-range order in excitons that couple neighboring $1D$ wires.
It would be interesting to analyze the fate of such a system of coupled Luttinger liquids at low energies. 
The phase fluctuations studied in this work mimic gauge fluctuations in one of the directions, a feature that generally applies to fluctuations in subsystem symmetry-broken states. 
While the microscopic Hamiltonian we considered only has density-density interactions between electrons in adjacent layers, one could consider correlated hopping terms as well, which also preserve the subsystem symmetry. Correlated hopping would enable a coupling between the fermions and the out-of-plane component of the electromagnetic vector potential. In particular, a non-zero AC conductivity is expected~\cite{lake2023nfl}. Additionally, we anticipate orbital coupling to in-plane magnetic fields, which can give rise to quantum oscillations that depend sensitively on the orientation of the magnetic field.

\textit{Acknowledgements}: We thank Amogh Anakru, Byungmin Kang, Ethan Lake, Patrick Lee, Allan MacDonald, Rahul Nandkishore, Adarsh Patri, Vladislav Poliakov, Andrew Potter, Zhengyan Darius Shi and Senthil Todadri for insightful discussions. We thank the anonymous referees for their constructive criticism.
A.K.\ was supported by the Gordon and Betty Moore Foundation EPiQS Initiative through Grant No.\ GBMF8684 at the Massachusetts Institute of Technology.
We thank the developers of the programming languages Julia \cite{Julia-2017} and Numbat \cite{Peter_Numbat_2024}, which were utilized for numerical computations.

\bibliography{references.bib}

\appendix

\onecolumngrid
\newpage
\section{Appendix A: Calculation of self-energy}
In this section, we provide details of the fermion self-energy calculation.
\begin{figure}[h]
    \centering
    \includegraphics[width=0.48\textwidth]{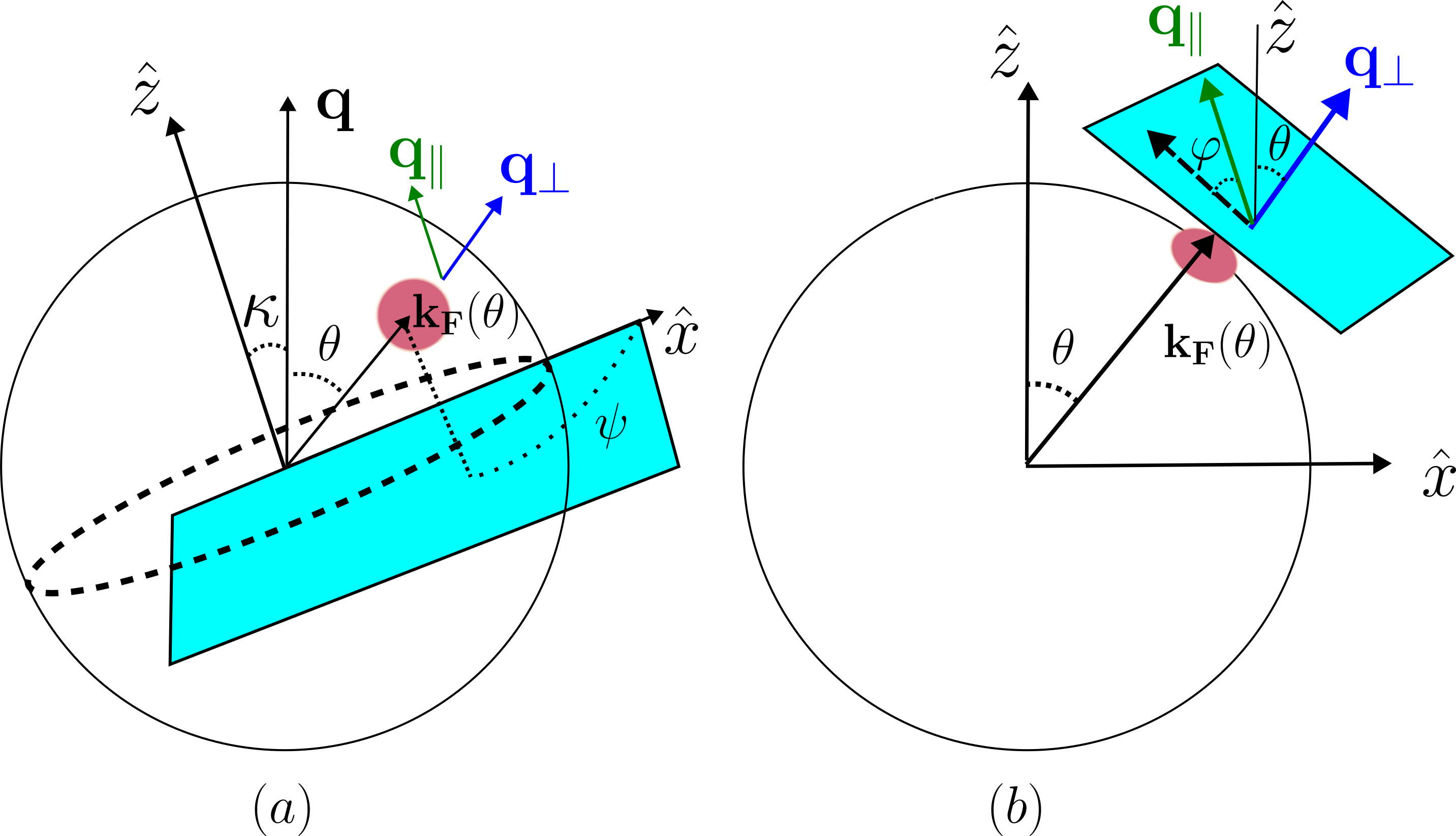}
    \caption{(a)We have to average the Boson self-energy contribution due to each Fermi surface patch (red patch) to obtain $\Pi^B_{\text{av}}$ in Eq.\eqref{eq:avg-fermi-surface-patch}. (b) Figure for calculating the momentum integrals of Fermion self-energy, Eq.\eqref{eq:Fermion-self-energy}. It follows that $\hat{q}_\pl \cdot \hat z = \cos\varphi \sin\theta$.}
    \label{fig:FS-patch}
\end{figure}
The boson self-energy contribution from a fermion patch labelled by $\theta$ is \cite{sachdev1999quantum},
\begin{align}
    \Pi^B_{\text{patch}}(\Omega, \v q) = \frac{2\Lambda k_{F,z}(\theta)^2 \Delta_0^2 m}{\pi v_F}\frac{|\Omega|}{|q_{\pl}|}
\end{align}
Defining angle $\kappa$ between $\v q$ and the $z$ axis (Fig.\ref{fig:FS-patch}(a)), and averaging over all the Fermi surface patches:
\begin{align}\label{eq:avg-fermi-surface-patch}
    \Pi^B_{\text{av}}(\Omega, \v q) =  \frac{1}{4\pi}\int d\psi d\theta \sin \theta \frac{2\Lambda k_{F,z}(\theta)^2 \Delta_0^2 m}{\pi v_F}\frac{|\Omega|}{q \sin \theta} \(\sin \kappa \sin \theta \cos \psi + \cos \kappa \cos \theta\)^2
\end{align}
yields the result quoted in the main text. In this calculation, we assumed that the dispersion is isotropic for simplicity, as it should not modify the universal scaling properties.

Now let us calculate the Fermion self-energy.
It is evident from Fig.\ref{fig:FS-patch}(b) that, $\hat{q}_{\parallel}\cdot \hat{z} = \cos \varphi \sin\theta$. 
Therefore, $1 + (\hat{q}_{\parallel}\cdot \hat{z})^2 =  1 + (\cos \varphi \sin\theta)^2 = \cos^2\varphi (2-\cos^2 \theta) + \sin^2 \varphi$, and, $(1 - (\hat{q}_{\parallel}\cdot \hat{z})^2) = \cos^2 \varphi \cos^2 \theta + \sin^2 \varphi$.
We first evaluate the $q_{\perp}$ integral to obtain
\begin{align}\label{eq:Fermion-self-energy}
    \Sigma(\omega_n) = &i \frac{4\Delta_0^2 k^2_{F,z}}{v_F} \int \frac{q_{\pl}d q_{\pl} d\varphi}{(2\pi)^2} \int \frac{d \Omega}{2\pi} \frac{\text{sgn}(\omega+\Omega)q_{\pl}}{\rho_s q_\pl^3 \(\cos^2 \varphi \cos^2 \theta + \sin^2 \varphi\) + \gamma |\Omega| \(\cos^2 \varphi (2-\cos^2 \theta) + \sin^2 \varphi\)} \nonumber \\
    &= i\frac{\Delta_0^2 k^2_{F,z}}{\pi^3 v_F \gamma}\text{sgn}(\omega_n) \int dq_{\pl} d\varphi \frac{q_\pl^2}{\cos^2 \varphi (2-\cos^2 \theta) + \sin^2 \varphi} \log \( \frac{\gamma |\omega_n| \frac{\cos^2 \varphi (2-\cos^2 \theta) + \sin^2 \varphi}{\cos^2 \theta \cos^2 \varphi + \sin^2 \varphi} + \rho_s q_\pl^3}{\rho_s q_{\pl}^3}\) \nonumber \\
    &= i\frac{\Delta_0^2 k^2_{F,z}}{3\pi^3 v_F \rho_s} \text{sgn}(\omega_n) |\omega_n| \log \left(\frac{\rho_s \Lambda^3}{\gamma |\omega_n|}\right) \int_0^{2\pi} d\varphi \frac{1}{\cos^2 \theta \cos^2 \varphi + \sin^2 \varphi} + \text{terms linear in }\omega_n\\
    &= i \frac{2\Delta_0^2 k_F^2 |\cos\theta|}{3\pi^2\rho_s v_F} \text{sgn}(\omega_n)|\omega_n| \log\left(\frac{\rho_s \Lambda^3}{\gamma |\omega_n|}\right) + \text{terms linear in }\omega_n.
\end{align}
As the imaginary part of the Matsubara self-energy is odd in $\omega_n$, it may seem that the lifetime may become negative. However, we need to calculate the imaginary part of the retarded self-energy $\Sigma^R(\omega) = \Sigma(\omega_n \rightarrow \frac{\omega + i 0^+}{i})$ to get the inverse-lifetime. Under analytical continuation, we obtain, $\text{sgn}(\omega_n) = \text{sgn}(\text{Im}[i \omega_n]) \rightarrow \text{sgn}(\text{Im}[\omega + i 0^+]) =1$. Therefore, the lifetime becomes an even function of the quasiparticle energy, as necessary to prevent spontaneous proliferation of quasiparticles and quasiholes, which would destroy the presumed ground state.
We obtain a quasiparticle lifetime $\tau(\omega) = \frac{1}{2}\text{Im}\left[ \Sigma^R(\omega)\right] \sim \frac{1}{|\omega| \log(1/|\omega|)}$.

\section{Appendix B: Calculation of specific heat}
\label{app:heat}
To calculate the specific heat in the low-temperature limit, we need to isolate the temperature-dependent part of the free energy integral, by subtracting the zero temperature free energy. This is similar to the analysis in Ref.~\cite{senthil2004weak},
\begin{equation}
    \delta F (T) = F(T) - F(0) =
    \frac{T}{2} \sum_{\Omega_n}\int \frac{d^3 q}{(2\pi)^3} \log \( \rho_s(q_x^2 + q_y^2) + \gamma \frac{|\Omega_n|}{q}\) - \frac{1}{2} \int \frac{d\Omega}{2\pi}\int \frac{d^3 q}{(2\pi)^3} \log \( \rho_s(q_x^2 + q_y^2) + \gamma \frac{|\Omega|}{q}\)
\end{equation}
We note that $\log(\rho_s (q_x^2 + q_y^2) + \gamma \frac{|\Omega|}{q}) \approx \log (\gamma\Lambda_1/q) - \int_0^{\Lambda_1} d\lambda \frac{\frac{\gamma}{q}}{\rho_s q_{2D}^2 + \gamma \frac{|\Omega|}{q} + \gamma \frac{\lambda}{q}}$ for large $\Lambda_1$. Since the $\log\Lambda_1$ term contributes a $T$-linear term to the free energy, it does not affect the specific heat $C =- T\frac{\partial^2 F(T)}{\partial T^2}$. Thus, we can drop it for our purpose, and take the limit $\Lambda_1 \rightarrow \infty$.

We also make use of the following identity, which holds for any function $h(\Omega_n)$ summed over Bosonic Matsubara frequencies. It can be proved by Poisson resummation.
\begin{equation}
    T\sum_{\Omega_m} h(\Omega_m) = \int_{-\infty}^\infty \frac{d\Omega}{2\pi} \sum_{n=-\infty}^\infty e^{\frac{i n\Omega }{T}} h(\Omega)
\end{equation}
Therefore,
\begin{equation}
\begin{aligned}
    \delta F (T) &= -\frac{1}{2} \int_{-\infty}^\infty \frac{d\Omega}{2\pi}\int \frac{d^3 q}{(2\pi)^3} \sum_{n\neq 0} e^{i n\Omega /T} \int_0^{\infty} d\lambda \frac{\gamma}{\rho_s q q_{2D}^2 + \gamma {|\Omega|} + \gamma {\lambda}}\\
    &= -\frac{1}{2} \int_{-\infty}^\infty \frac{d\Omega}{2\pi}\int \frac{d^3 q}{(2\pi)^3} \sum_{n\neq 0} e^{i n\Omega /T} \int_0^{\infty} d\lambda \int_0^\infty du {\gamma} e^{-u (\rho_s q q_{2D}^2 + \gamma {|\Omega|} + \gamma {\lambda})}.
\end{aligned}
\end{equation}
In the last step we introduced yet another dummy variable $u$, so that we can next interchange the order of the integrations to eventually isolate the dominant term in the free energy. We first integrate w.r.t.\ $\Omega$ and obtain,
\begin{align}
\delta F(T)= -\frac{1}{2} \int \frac{d^3 q}{(2\pi)^3} \int_0^{\infty} d\lambda \int_0^\infty du {\gamma} \underbrace{\sum_{n\neq 0} \frac{1}{u^2 \gamma^2 T^2 + n^2}}_{=\frac{\pi u \gamma T \coth(\pi u \gamma T) - 1}{u^2\gamma^2T^2}} \frac{u \gamma T^2}{\pi} e^{-u (\rho_s q q_{2D}^2 + \gamma {\lambda})}
\end{align}
Now, let's do the $\lambda$ integral,
\begin{align}
\delta F(T)&= -\frac{1}{2} \int \frac{d^3 q}{(2\pi)^3} \int_0^\infty du {\gamma} e^{-u (\rho_s q q_{2D}^2)} \frac{T^2}{\pi} \frac{\pi u \gamma T \coth(\pi u \gamma T) - 1}{u^2\gamma^2T^2}\\
&= -\frac{1}{2} \int_0^\Lambda \frac{q^2 dq}{(2\pi)^3}\int_0^\pi d\theta \sin\theta (2\pi) \int_0^\infty du {\gamma} e^{-u \rho_s q^3 \sin^2 \theta} \frac{T^2}{\pi} \frac{\pi u \gamma T \coth(\pi u \gamma T) - 1}{u^2\gamma^2T^2}
\end{align}
The $u$ integral can be broken into two pieces, one from $0$ to $\frac{1}{\pi \gamma T}$, and the other from $\frac{1}{\pi \gamma T}$ to $\infty$. In the first regime, $T^2 \frac{\pi u \gamma T \coth(\pi u \gamma T) - 1}{u^2\gamma^2T^2} \approx \frac{\pi^2}{3}T^2$, and in the second regime, $T^2 \frac{\pi u \gamma T \coth(\pi u \gamma T) - 1}{u^2\gamma^2T^2} \approx \frac{\pi T}{u\gamma}$, which is $T$-linear and does not contribute to the specific heat. Therefore it can be dropped for our purpose. After integrating over $u$, we obtain,
\begin{equation}   
\begin{aligned}
\delta F(T) &\approx -\frac{T^2 \pi^2}{6} \int_0^\Lambda \frac{2\pi q^2 dq}{(2\pi)^3}\int_0^\pi d\theta \sin \theta \frac{\gamma}{\pi} \frac{1 - e^{-\frac{\rho_s q^3 \sin^2 \theta}{\pi\gamma T}}}{\rho_s q^3 \sin^2 \theta}\\
&= -\frac{T^2 \gamma}{36 \pi \rho_s} \int_0^\frac{\pi}{2} \frac{d\theta}{\sin \theta} \int_0^\frac{\rho_s \Lambda^3 \sin^2\theta}{\pi \gamma T} dt \frac{1-e^{-t}}{t}
\end{aligned}
\end{equation}
We make use of the fact that $\int_0^\alpha dt \frac{1-e^{-t}}{t} \sim \log(\alpha)$ for $\alpha \gg 1$, but it is approximately equal to $\alpha$ for $\alpha \ll 1$. To get the leading scaling dependence as a function of temperature, we have to break the $\theta$ integral into one from $0$ to $\sqrt{\frac{\pi \gamma T}{\rho_S \Lambda^3}}$, and one from $\sqrt{\frac{\pi \gamma T}{\rho_S \Lambda^3}}$ to $\pi/2$. The first generates an overall $T^2$ dependence, but the latter dominates at low temperature. The leading term from the second integral is,
\begin{equation}   
\begin{aligned}
\delta F(T)
&\approx -\frac{T^2 \gamma}{36 \pi \rho_s} \int_{\sqrt{\frac{\pi\gamma T}{\rho_s \Lambda^3}}}^\frac{\pi}{2} \frac{d\theta}{\sin \theta} \log\left(\frac{\rho_s \Lambda^3 \sin^2\theta}{\pi \gamma T}\right)
&\approx \boxed{-\frac{T^2 \gamma}{144 \pi \rho_s} \left(\log\left(\frac{\rho_s \Lambda^3}{\pi \gamma T}\right)\right)^2}
\end{aligned}
\end{equation}
which results in $\boxed{C \sim T(\log (1/T))^2}$. Being proportional to $\gamma \sim \Delta_0^2$, this modification of specific heat would vanish when the subsystem symmetry breaking condensate ceases to exist.

In the standard case of an isotropic Fermi surface coupled to a gauge field, the free energy behaves as
$\delta F(T) \sim T^2 \int dq \, \frac{1 - \text{exp}\(-\rho_s q^3/(\pi \gamma T)\)}{q}$.
This expression exhibits an IR divergence in the $q$-integral, which is regulated by the temperature, yielding $F(T) \sim -T^2 \log(1/T)$.
In the present scenario, there is an additional factor of $\sin \theta$ in the denominator due to absence of dispersion along $z$, leading to an extra source of logarithmic divergence.

\section{Appendix C: Modification to transport properties due to Fermion-Goldstone mode coupling}
\label{app:transport}
Let us first consider the case where $\phi$ is not Landau-damped and behaves as a propagating field, meaning the probing energy scale is above the Landau-damping threshold. The transport scattering rate, $1/\tau_{\text{tr}}$, can be derived by incorporating a weighting factor of $1 - \cos \theta_{\v q}$, where $\theta_{\v q}$ represents the change in the electron's momentum angle during a scattering event with momentum transfer $\v q$. Using this factor in the Fermi golden rule expression for the inverse lifetime, the transport scattering rate for an electron at energy $\epsilon >0$ relative to the chemical potential, at patch $\theta$, is given by~\cite{lee1992gauge}:
\begin{align}\label{eq:app:transport-scattering-rate}
    \frac{1}{\tau_{\text{tr}}(\eps)} =& 8\Delta_0^2 k_{F,z}(\theta)^2 \int_0^{\infty} d \Omega \int_{\v q}(1-\cos \theta_{\v q}) \Theta(\eps_{\v k'}) \delta(\eps - \eps_{\v k'} - \Omega) \text{Im} D^{\text{ret}}(\v q, \Omega)
\end{align}
where $D^{\text{ret}}$ is the retarded propagator of $\phi$. For small momentum transfer, we can approximate $1-\cos \theta_{\v q} \approx \frac{1}{2}(\frac{q}{k_F})^2$. When $\phi$ is a propagating field, 
\begin{align}
D^{\text{ret}}(\v q, \Omega) = \frac{1}{\rho_s}\frac{1}{\Omega/c - q_{\pl}\sqrt{1-(\hat{q}_{\parallel} \cdot \hat{z})^2} + i\delta}\frac{1}{\Omega/c + q_{\pl}\sqrt{1-(\hat{q}_{\parallel} \cdot \hat{z})^2} + i\delta},
\end{align}
where $\Omega$ is a real frequency and $\delta$ is a positive infinitesimal broadening term.

By decomposing the $\v q$ integral into normal ($q_{\perp}$) and tangential ($q_{\parallel}$) to the Fermi surface, and linearizing the fermionic dispersion: $\eps_{\v k'} \approx v_F k'_{\perp}$, we obtain
\begin{align}
    \frac{1}{\tau_{\text{tr}}(\eps)} &= \frac{2\Delta_0^2 k_{F,z}^2 c}{\rho_s v_F k_F^2} \int_0^{\eps}\frac{d\Omega}{\Omega} \int \frac{d^2 q_{\parallel}}{(2\pi)^2} q_{\parallel}^2 \delta\(q_{\parallel}\sqrt{1-(\hat{q}_{\parallel} \cdot \hat{z})^2} - \frac{\Omega}{c}\) \nonumber \\
    &=\frac{\Delta_0^2 (1+\cos^2 \theta)}{6\pi \rho_s c^2 v_F  |\cos \theta|} \eps^3
\end{align}

The total conductivity summed over all patches, results in a resistivity $\rho \sim T^3$ at finite temperature $T$. 
By contrast, electron-phonon scattering in a $3D$ metal is known to result in a smaller $\rho \sim T^5$. The two additional powers in temperature can be attributed to the electron-phonon coupling vertex scaling linearly with the phonon momentum $q$ for small $q$, whereas the electron-phase fluctuation coupling is a $q$-independent constant for small $q$. 
We will now conduct a similar analysis for electron scattering off Landau-overdamped phase fluctuations.
$\text{Im}D^{\text{ret}}(\v q, \Omega)$ for the Landau-overdamped phase fluctuations is
\begin{align}
    \text{Im}D^{\text{ret}}(\v q_{\pl}, \Omega) = \frac{1}{\rho_s} \frac{\gamma \Omega q_{\pl}}{\gamma^2 \Omega^2 + q_\pl^2 \(q_{\pl,x}^2 \cos^2 \theta + q_{\pl,y}^2\)}
\end{align}
Here, $q_{x}$
We have omitted the unimportant anisotropy factor in the boson self-energy, consistent with the approach taken in the specific heat analysis.
We then have
\begin{align}
    \frac{1}{\tau_{\text{tr}}(\eps)} &= 8 \Delta_0^2 k_{F,z}(\theta)^2 \int_0^{\infty} d\Omega \int \frac{d^2 q_{\pl}}{(2\pi)^2}\frac{dq_{\perp}}{2\pi} \frac{q_{\pl}^2}{2k_F^2}\t\Theta(v_F q_{\perp}) \delta(\eps - v_F q_{\perp} - \Omega) \text{Im} D^{\text{ret}}(\v q_{\pl}, \Omega) \nonumber \\
    &= \frac{\Delta_0^2 \gamma \cos^2 \theta}{2\pi^3\rho_s v_F} \int_0^{\eps} d\Omega \Omega \int_0^{2\pi} \frac{d\varphi}{\(\cos^2 \theta \cos^2 \varphi + \sin^2 \varphi\)^2} \int_0^{\infty} dq_{\pl} \frac{q_{\pl}^4}{q_\pl^6 + \gamma^2 \Omega^2/\(\cos^2 \theta \cos^2 \varphi + \sin^2 \varphi\)^2} \nonumber \\
    &= \frac{\Delta_0^2 \gamma^{2/3} \cos^2 \theta}{10 \pi^2 \rho_s v_F} \eps^{5/3} g(\theta)
\end{align}
where $g(\theta) \equiv \int_0^{2\pi} d\varphi \(\cos^2 \theta \cos^2 \varphi + \sin^2 \varphi\)^{-5/3} \sim \cos^{-7/3} \theta$ for $\theta \rightarrow \pi/2$. Combining with $\cos^{2} \theta$ factor, we find that $1/\tau_{\text{tr}}(\eps) \sim \eps^{5/3}/|\cos^{1/3} \theta|$ for $\theta \approx \pi/2$.
Summing over all patches results in a resistivity $\rho \sim T^{5/3}$. 

\section{Appendix D: Effect of inter-layer hopping on lifetime of quasiparticles, and its temperature dependence}
In this appendix we study the effect of inter-layer hopping on the non-Fermi liquid properties. Here we show that an interlayer hopping term with amplitude $t_e$ generates Fermi liquid behavior in excitations of energy $|\epsilon| < t_e$, but non-Fermi liquid behavior otherwise.

We estimate the lifetime of an excitation with energy $\epsilon > 0$ relative to the chemical potential, in a manner similar to the one in the previous section. As we are estimating lifetime of an excitation and not the transport scattering timescale, we exclude the $1-\cos\theta_{\v q}$ factor that was present in Eq.\eqref{eq:app:transport-scattering-rate},
\begin{align}\label{eq:app:transport-scattering-rate}
    \frac{1}{\tau_{\rm lifetime}(\eps)} =& 8\Delta_0^2 k_{F,z}(\theta)^2 \int_0^{\infty} d \Omega \int_{\v q} \theta(\eps_{\v k'}) \delta(\eps - \eps_{\v k'} - \Omega) \text{Im} D^{\text{ret}}(\v q, \Omega),
\end{align}
where $D^{\text{ret}}(\v q, \Omega)$ is the retarded propagator of $\phi$, which acquires a mass $m_e \sim t_e/c^2$ when the hopping amplitude $t_e \neq 0$. In particular,
\begin{align}
D^{\text{ret}}(\v q, \Omega) = \frac{1}{\rho_s}\frac{1}{\Omega/c - \sqrt{q_{\pl}^2 (1-(\hat{q}_{\parallel} \cdot \hat{z})^2) + m_e^2 c^2} + i\delta}\frac{1}{\Omega/c + \sqrt{q_{\pl}^2 (1-(\hat{q}_{\parallel} \cdot \hat{z})^2) + m_e^2 c^2} + i\delta},
\end{align}
where $\Omega$ is a real frequency and $\delta$ is a positive infinitesimal broadening term. Integrating over $q_\perp$ we obtain,
\begin{align}
    \frac{1}{\tau_{\rm lifetime}(\eps)} &= \frac{4\Delta_0^2 k_{F,z}^2 c}{\rho_s v_F} \int_0^{\eps}\frac{d\Omega}{\Omega} \int \frac{d^2 q_{\parallel}}{(2\pi)^2}  \delta\(\sqrt{q_{\parallel}^2\left(1-(\hat{q}_{\parallel} \cdot \hat{z})^2\right) + m_e^2 c^2} - \frac{\Omega}{c}\).
\end{align}
Clearly, if $\epsilon < m_e c^2$, the argument of the Dirac $\delta$ function is always positive and the contribution of Fermion-Boson scattering to the scattering rate $1/\tau$ is zero. Of course, quasiparticles will gain a finite lifetime ($1/\tau \sim \epsilon^2$) due to Coulomb repulsion, and such quasiparticles with energy $\epsilon < m_e c^2$ will behave just like those in a Fermi liquid. 

However, the quasiparticles with energy $\epsilon$ greater than the Bosonic gap $m_e c^2$ set by interlayer hopping will have a finite lifetime,
\begin{align}
    \frac{1}{\tau_{\rm lifetime}(\eps)} &= \frac{4\Delta_0^2 k_{F,z}^2 c}{\rho_s v_F} \int_0^{\eps}\frac{d\Omega}{\Omega} \int \frac{d^2 q_{\parallel}}{(2\pi)^2}  \delta\(\sqrt{q_{\parallel}^2\left(1-(\hat{q}_{\parallel} \cdot \hat{z})^2\right) + m_e^2 c^2} - \frac{\Omega}{c}\) \nonumber\\
    &= \frac{2 \Delta_0^2 k_F^2   |\cos\theta| \Theta(\epsilon - m_e c^2)}{\pi \rho_s v_F} (\epsilon- m_e c^2)   ,
\end{align}
which roughly captures the behavior of the imaginary part of the self-energy upto the logarithmic factor.

Clearly, if the temperature $T$ is lower than the tunneling amplitude $t_e$, the typical energy of the Fermionic excitations will be less than $m_e c^2$, and they will have lifetime $\tau \sim 1/\epsilon^2$ due to Coulomb scattering, similar to that in regular Fermi liquids. In contrast, when the temperature is greater than $t_e$, quasiparticle states with energy greater than $m_e c^2$ will be thermally accessible and will have a much shorter lifetime ($\tau \sim 1/\epsilon$ --- more precisely, $\tau \sim 1/(\epsilon \log (1/\epsilon))$, see Eq.\eqref{eq:fermion-self-energy-result}), just like that in a marginal Fermi liquid.

When the hopping amplitude $t_e$ is exponentially suppressed, the effect of the very low energy states (that behave like Fermi liquid) will be washed away, and the system will behave like a marginal Fermi liquid when temperature $T \gtrsim t_e$. When $t_e$ is exponentially small, only the marginal Fermi liquid behavior will show up in experimentally attainable low temperatures.

\end{document}